\documentclass[10pt,letterpaper]{article}
\usepackage[top=0.85in,left=0.65in,footskip=0.75in,marginparwidth=2in]{geometry}

\usepackage[utf8]{inputenc}
\usepackage{amsmath}

\usepackage{cite}

\usepackage{nameref,hyperref}

\usepackage[right]{lineno}

\usepackage{microtype}
\DisableLigatures[f]{encoding = *, family = * }

\usepackage{changepage}

\usepackage[aboveskip=1pt,labelfont=bf,labelsep=period,singlelinecheck=off]{caption}

\makeatletter
\renewcommand{\@biblabel}[1]{\quad#1.}
\makeatother

\usepackage{lastpage,fancyhdr,graphicx}
\usepackage{epstopdf}
\fancyheadoffset[L]{2.25in}
\fancyfootoffset[L]{2.25in}

\usepackage{color}

\definecolor{Gray}{gray}{.25}

\usepackage{graphicx}

\usepackage{sidecap}

\usepackage{wrapfig}
\usepackage[pscoord]{eso-pic}
\usepackage[fulladjust]{marginnote}
\reversemarginpar

\begin{document}
\vspace*{0.35in}

\begin{flushleft}
{\Large
\textbf\newline{PERFORMANCE OF DIFFERENTIAL PROTECTION APPLIED TO COLLECTOR CABLES OF OFFSHORE WIND FARMS WITH MMC-HVDC TRANSMISSION}
}
\newline
\\
Moisés J. B. B. Davi\textsuperscript{1},
Felipe V. Lopes\textsuperscript{2},
Vinícius A. Lacerda\textsuperscript{3},
Mário Oleskovicz\textsuperscript{1},
Oriol Gomis-Bellmunt\textsuperscript{3}
\\
\bigskip
\bf{1} University of São Paulo, Department of Electrical Engineering, São Carlos, Brazil
\\
\bf{2} Federal University of Paraíba, Department of Electrical Engineering, João Pessoa, Brazil
\\
\bf{3} Universitat Politècnica de Catalunya, Barcelona, Spain
\\
\bigskip
* moisesdavi@usp.br

\end{flushleft}

\section*{Abstract}
The ongoing global transition towards low-carbon energy has propelled the integration of offshore wind farms, which, when combined with Modular Multilevel Converter-based High-Voltage Direct Current (MMC-HVDC) transmission, present unique challenges for power system protection. In collector cables connecting wind turbines to offshore MMC, both ends are supplied by Inverter-Based Resources (IBRs), which modify the magnitude and characteristics of fault currents. In this context, this paper investigates the limitations of conventional differential protection schemes under such conditions and compares them with enhanced strategies that account for sequence components. Using electromagnetic transient simulations of a representative offshore wind farm modeled in PSCAD/EMTDC software, internal and external fault scenarios are assessed, varying fault types and resistances. The comparative evaluation provides insights into the sensitivity and selectivity of differential protection and guides a deeper conceptual understanding of the evolving protection challenges inherent to future converter-dominated grids.


\section{Introduction}

In recent decades, wind farms have become an increasingly important part of the global power system, driven by the energy transition and the growing adoption of renewable energy resources \cite{LEE2025}. These developments have led to the need for robust and reliable protection solutions tailored to the specific characteristics of Inverter-Based Resources (IBRs), which affect fault current behavior \cite{LOPES2024}. In this context, for onshore wind farms, there is broad acceptance among professionals regarding the effectiveness of communication-assisted protection functions, such as the differential protection, for interconnection lines. Such schemes typically offer high sensitivity and selectivity, even in complex fault scenarios involving IBRs \cite{LOPES2024}.

Nevertheless, the expansion of offshore wind generation and the use of Modular Multilevel Converter-based High-Voltage Direct Current (MMC-HVDC) transmission systems introduce new challenges for the protection of collector cables connecting Offshore MMC (OMMC) to wind turbine converters \cite{WILL2025}. In these sections, both line terminals are supplied by IBRs, whose control strategies not only actively limit the magnitude of fault currents but also alter their characteristics, making them distinct from the conventional fault currents for which traditional protections were originally designed \cite{CHOW2021A,BINI2021,KASZTENNY2021}. 

Such changes in fault current behavior can impact key protection algorithm quantities, such as operating and restraint ones of conventional differential schemes, making it more difficult to discriminate internal and external faults reliably. As a result, traditional differential protection strategies may have insufficient sensitivity, which can compromise fault detection and potentially leave these important transmission paths unprotected during critical events \cite{CHAO2024}. Although this is a current and relevant topic, it is still underexplored in the literature. The existing works in this scope, such as \cite{CHAO2024,GOKSU2017,ZHENG2022}, assess only the limitations of phase differential protection units, and in a restricted way, that is, by exploring fault scenarios that are not sufficiently representative for all fault types and resistance levels.
	
In this context, the present paper investigates the performance of different strategies for differential protection applied to offshore wind farms, considering faults in the collector cables between wind generators and OMMC. A wide range of fault scenarios was simulated using detailed modeling in PSCAD/EMTDC of a realistic offshore wind farm, including both internal and external faults on the collector cables, besides covering multiple fault types (single-phase, two-phase, two-phase-to-ground, and three-phase faults) and a diverse set of fault resistances. The study compares the sensitivity and selectivity of conventional phase current-based differential protections with approaches based on negative- and zero-sequence components.  

The results evidence that even differential protections, known for their sensitivity and selectivity, face application challenges in collector cables of offshore wind farms. The findings are intended to support not only protection engineers and system operators but also future research aiming at the high-reliability protection of converter-dominated power grids. 

\section{Assessed Differential Protections}

Differential protection selectively safeguards equipment or line segments, with the protected zone set by current transformers (CTs) at both terminals \cite{BLACK2014}. During normal operation or external faults, the terminal currents entering and leaving the zone (e.g., $\vec{I}_\text{local}$ and $\vec{I}_\text{remote}$) are ideally equal in magnitude and opposite in direction, resulting in a net sum of zero and preventing relay operation. An internal fault disrupts this balance, yielding a nonzero differential current that can trip the relay if it exceeds a predefined threshold. However, CT errors, especially during high through-current external faults, can lead to measurement inaccuracies and unwanted relay operation if not properly mitigated \cite{BLACK2014}. 

To address these challenges, percentage (or restraint) differential schemes are implemented. These incorporate both an operator and a restraint element within the relay logic. The operating quantity ($I_\text{OP}$) is typically derived as the magnitude of the vector sum of the local and remote CT current phasors \cite{BLACK2014}:

\begin{equation}
	I_\text{OP} = \left|\vec{I}_\text{local} + \vec{I}_\text{remote}\right|
\end{equation}

The restraint quantity ($I_\text{RST}$), on the other hand, is defined as the sum of the magnitudes of the local and remote terminal currents \cite{BLACK2014}:
\begin{equation}
	I_\text{RST} = |\vec{I}_\text{local}| + |\vec{I}_\text{remote}|
\end{equation}
The relay is set to operate only if the operating current exceeds a threshold that is a function of the restraint current \cite{BLACK2014}:

\begin{equation}
	I_\text{OP} > K \cdot I_\text{RST} + K_0
\end{equation}

where $K$ is the restraint slope, and $K_0$ is a constant representing the minimum pick-up threshold. For the studies in this paper, $K$ was assumed to be 50\% and $K_0$ was set to 0.3 p.u. of the nominal current.

Table~\ref{tab:differential_quantities} summarizes the specific formulations for the operating and restraint quantities applied in phase (87L), negative-sequence (87Q), and zero-sequence (87G) percentage differential protections. While the core principle of the percentage differential function, that is, the comparison between the operate and restraint quantities, remains consistent across all schemes, the choice of the measured current component (phases ($\vec{I}_{A}$, $\vec{I}_{B}$, and $\vec{I}_{C}$), negative-sequence ($\vec{I}_{2}$), or zero-sequence ($\vec{I}_{0}$)) tailors each protection method to address different fault types. 

\begin{table}[!b]
	\centering
	\renewcommand{\arraystretch}{1.2}	
	\caption{\centering $I_\text{OP}$ and $I_\text{RST}$ for each differential protection.}
	\label{tab:differential_quantities}	
	\begin{tabular*}{\linewidth}{@{\extracolsep{\fill}}ccc}
		\hline
		\textbf{Type} & \textbf{$I_\mathrm{OP}$} & \textbf{$I_\mathrm{RST}$} \\
		\hline
		$87L_{a,b,c}$ & $|\vec{I}_{\rm loc-a,b,c} + \vec{I}_{\rm rem-a,b,c}|$ & $|\vec{I}_{\rm loc-a,b,c}| + |\vec{I}_{\rm rem-a,b,c}|$ \\
		$87Q$ & $|\vec{I}_{2,\rm loc} + \vec{I}_{2,\rm rem}|$ & $|\vec{I}_{2,\rm loc}| + |\vec{I}_{2,\rm rem}|$ \\
		$87G$ & $|\vec{I}_{0,\rm loc} + \vec{I}_{0,\rm rem}|$ & $|\vec{I}_{0,\rm loc}| + |\vec{I}_{0,\rm rem}|$ \\
		\hline
	\end{tabular*}
\end{table}

\section{Test System}

\begin{figure*}[!t]
	\centering\includegraphics[width=1\linewidth]{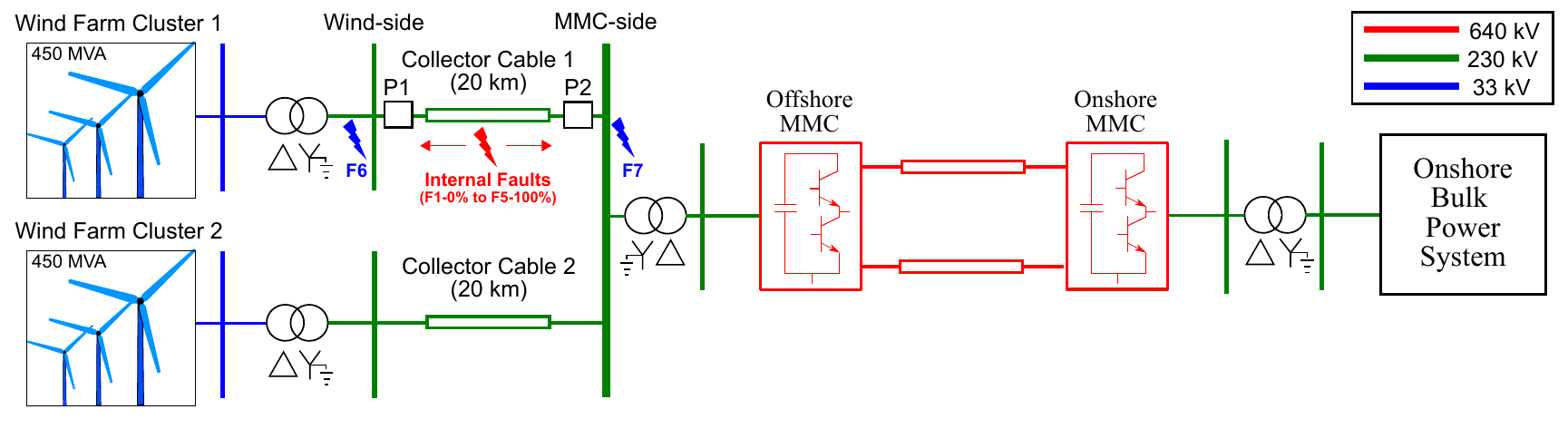}
	\caption{Test system.}\label{fig:Fig_1}
\end{figure*}

To evaluate the performance of protection schemes under the described conditions, a detailed model of a realistic offshore wind farm interconnected via MMC-HVDC was developed using PSCAD/EMTDC, as depicted in Fig. \ref{fig:Fig_1}. The modeled offshore wind farm consists of two wind turbine clusters, each with an installed capacity of 450 MVA. Both clusters are equipped with full-converter (type 4) wind turbines, linked to the OMMC via 20 km, 230 kV collector cables. For the purpose of this study, the collector cable associated with Cluster 1 was selected for analysis.

Measurement points P1 and P2 (see Fig. \ref{fig:Fig_1}) are considered at both terminals of the collector cable under investigation, and the analysis encompasses both internal faults along the collector cable 1 (F1, F2, F3, F4, and F5, corresponding to faults at 0\%, 25\%, 50\%, 75\%, and 100\% of the line length, respectively), and external faults (denoted as F6 for a fault at the 230 kV side of the transformer connected to Cluster 1 and as F7 for a fault at the 230 kV side of the transformer on the OMMC side). The fault scenarios included multiple fault types (AG, AB, ABG, and ABC) and resistances (varying the phase-to-phase resistance between 0, 2.5, 5, and 10 $\Omega$, and the phase-to-ground resistance between 0, 10, 25, and 50 $\Omega$). 

\begin{figure}[!b]
	\centering
	\includegraphics[width=0.65\textwidth]{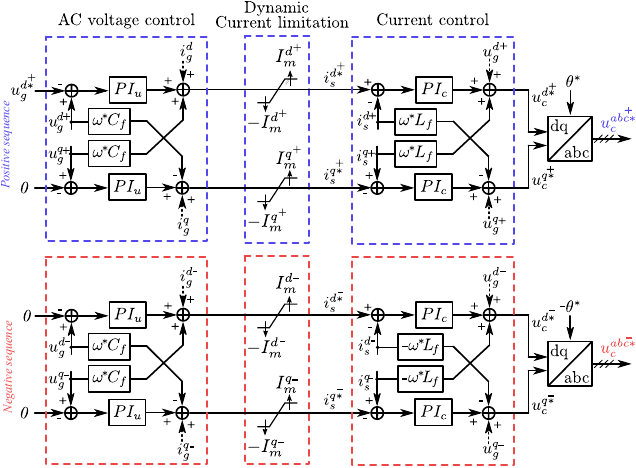}

	\caption{GFM voltage and current controls.}
	\label{fig_GFOR_innerloops}

\end{figure}

Concerning the IBRs' control strategies, the OMMC adopts an AC voltage-frequency control approach to provide the voltage reference needed for the operation of the wind turbines \cite{REF21}. The voltage control is implemented using the cascaded configuration, shown in Fig.~\ref{fig_GFOR_innerloops} for the positive and negative sequences. In fault situations, if the positive-sequence current exceeds a preset threshold, the OMMC limits the amplitude to 1.1 p.u. \cite{REF22}. 

For the negative-sequence component control loop, two scenarios are evaluated. In the first case, labeled as Control (C1), the control loop actively suppresses negative-sequence current during faults \cite{CHAO2024}. In the second scenario, referred to as Control 2 (C2), there is no forced suppression of negative-sequence currents during faults; on the contrary, this component is prioritized under such circumstances.

With regard to the control of the wind turbines, strategies that prioritize the injection of positive-sequence reactive current during faults have been adopted \cite{REF24}. However, recognizing the existence of other control strategies, the discussions to be conducted will also explore the implications for controls that provide negative-sequence reactive current \cite{IEEE2022}. 

\section{Results and Discussions}

In this section, the performance results of the differential protections 87L ($87L_{a}$, $87L_{b}$, and $87L_{c}$), 87Q, and 87G are presented for internal and external faults on the evaluated collector cable 1. For each differential element and fault type, four fault resistance levels were assessed, referenced as detailed in Table \ref{tab:fault_resistances}. Moreover, in order to provide a representative analysis, the differential planes ($I_\text{RST}$ vs. $I_\text{OP}$) were illustrated, indicating the convergence point of the differential protection elements 50 ms after the fault inception.

\begin{table}[!b]
	\centering
	\renewcommand{\arraystretch}{1.2}
	\caption{\centering Fault resistance values associated with R1--R4 for \\ each fault type.}
	\begin{tabular}{lcccc}
		\hline
		\textbf{Fault Type}           & \textbf{R1 ($\Omega$)} & \textbf{R2 ($\Omega$)} & \textbf{R3 ($\Omega$)} & \textbf{R4 ($\Omega$)} \\
		\hline
		AG ($R_{g}$)   & 0     & 10     & 25     & 50    \\
		AB ($R_{ph}$)      & 0     & 2.5    & 5      & 10    \\		
		ABG ($R_{g}$/$R_{ph}$)          & 0/0   & 2.5/10 & 5/25   & 10/50 \\
		ABC ($R_{ph}$)                    & 0     & 2.5    & 5      & 10    \\
		\hline
	\end{tabular}
	\label{tab:fault_resistances}
\end{table}

\subsection{Internal Phase-to-Ground (PG) Faults}

\begin{figure}[!b]

	\centering\includegraphics[width=0.45\linewidth]{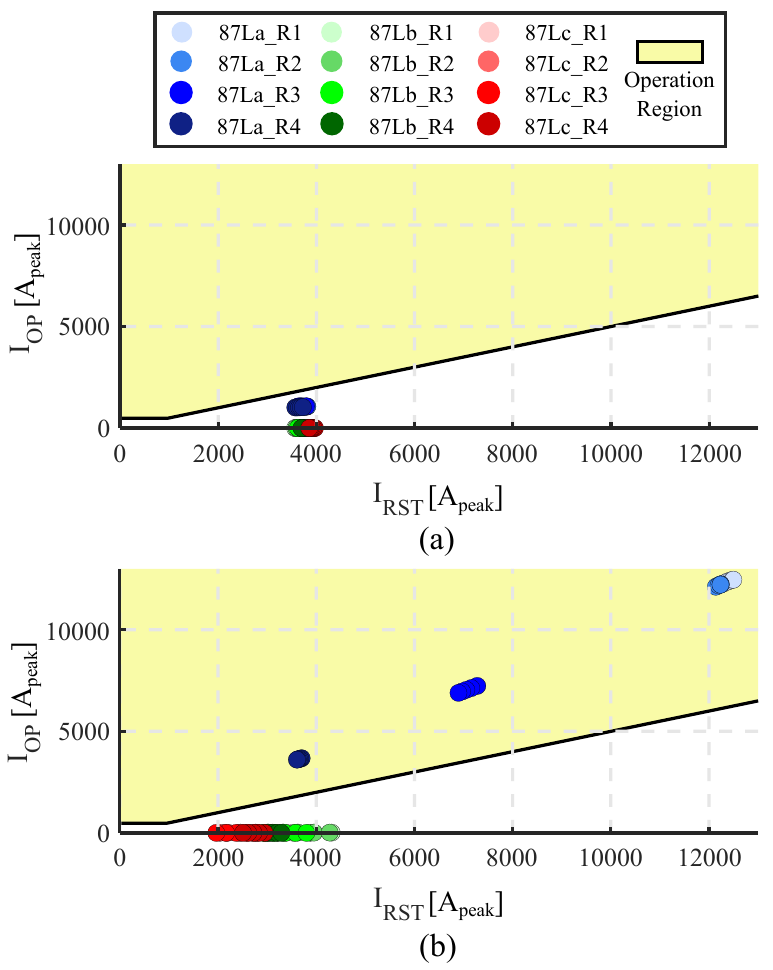}

	\caption{Responses of $87L_{a}$, $87L_{b}$, and $87L_{c}$ elements for AG faults, considering (a) C1 and (b) C2 controls.}\label{fig:Dif-PG-Fase}
\end{figure}

\begin{figure}[!b]

	\centering\includegraphics[width=0.45\linewidth]{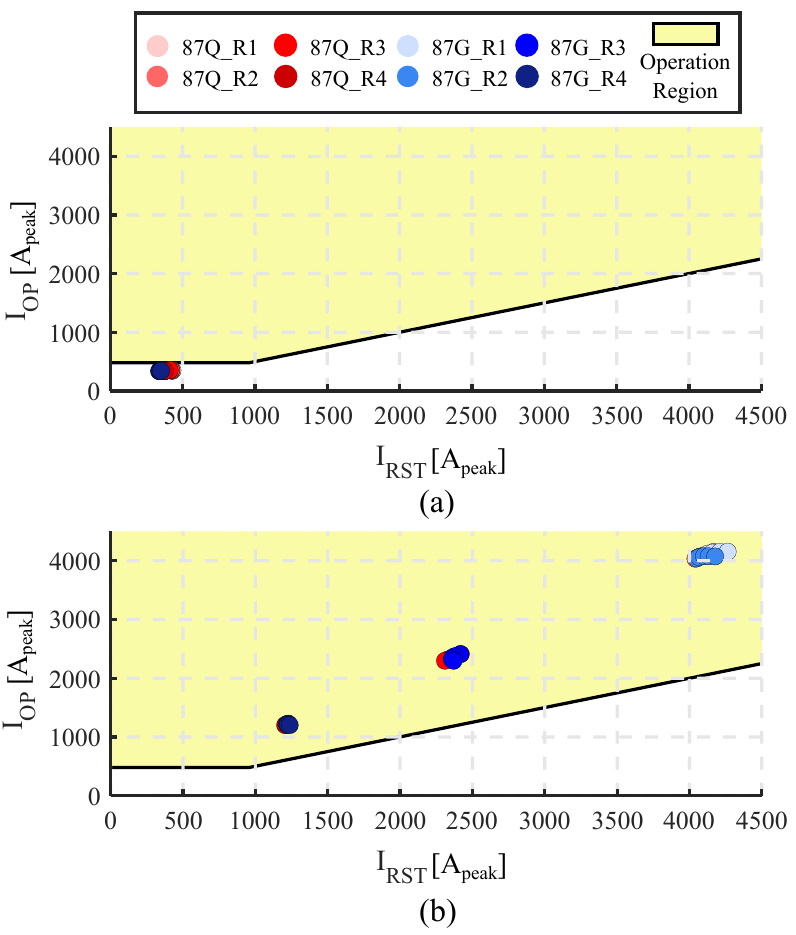}
	
	\caption{Responses of $87Q$ and $87G$ elements for AG faults, considering (a) C1 and (b) C2 controls.}\label{fig:Dif-PG-Seq}
\end{figure}

Figs. \ref{fig:Dif-PG-Fase} and \ref{fig:Dif-PG-Seq}  illustrate the convergence points of the phase and sequence differential elements, respectively. The results show that, when considering control C1, that is, the forced suppression of negative-sequence current in the OMMC, none of the differential elements exhibit sufficient sensitivity to trigger differential protection operation under internal faults.

\begin{figure}[!t]
	\centering\includegraphics[width=0.5\linewidth]{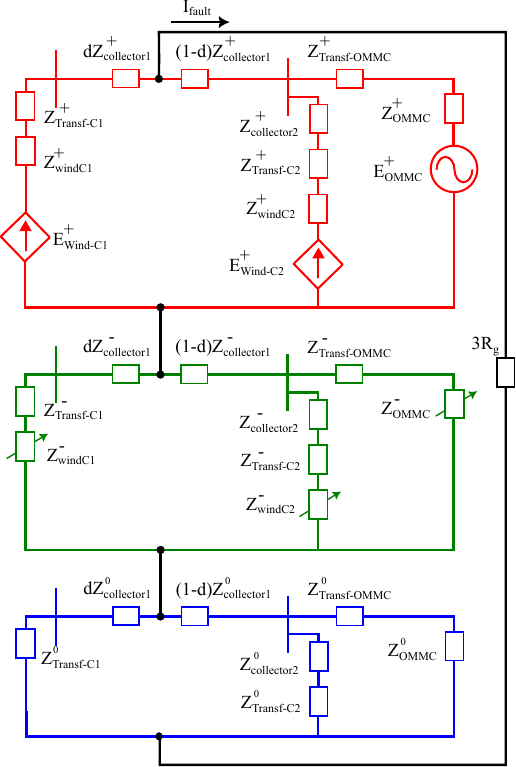}

	\caption{Sequence network for PG fault.}\label{fig:CircSeqFT}

\end{figure}

This condition can be explained by the sequence network diagram for PG faults, shown in Fig. \ref{fig:CircSeqFT}. In this figure, the fault distance ($d$), the impedances of collector cables 1 and 2 ($Z_{collector1}$ and $Z_{collector2}$), the impedance of the transformers of clusters 1, 2 and OMMC ($Z_{Transf-C1}$, $Z_{Transf-C2}$, $Z_{Transf-OMMC}$), as well as the equivalent series impedances of clusters 1, 2 and OMMC ($Z_{wind-C1}$, $Z_{wind-C2}$, $Z_{OMMC}$), for positive, negative, and zero sequence, are represented. As can be observed, when considering control C1 in the OMMC, since the sequence circuits are connected in series and the wind turbines also include the characteristic of negative-sequence current suppression, the fault current observed ($I_{fault}$) is practically null. In other words, under these conditions, the three parallel paths of the negative-sequence circuit are open circuits, suppressing the fault current and, consequently, compromising the operation of differential protections. 

It is worth mentioning that, in practice, residual levels of negative-sequence current may be measured, since the presence of filters and control characteristics makes negative-sequence current suppression non-ideal. Thus, when considering more sensitive differential elements such as 87Q and 87G (Fig. \ref{fig:Dif-PG-Seq}), the differential protection may operate if lower pickup values or even slopes below 50\% are considered. On the other hand, it should be considered that by making the differential protection more sensitive with its reparametrization, its selectivity for external faults to the protected zone may also be compromised. 

When considering control C2 in the OMMC, there is proper and sensitive operation of all evaluated differential elements. Similar situations could also be obtained when considering controls that provide negative-sequence current by the wind turbines \cite{IEEE2022}.

Another discussion point concerns the configuration of transformer windings. For the evaluated test system, since the transformers on both sides of the collector cable are configured as grounded wye, there is a possible path for fault current through the zero-sequence circuit. However, depending on these transformer connections, the zero-sequence circuit may also result in near-zero fault currents for PG faults.

\begin{figure}[!b]
	\centering\includegraphics[width=0.45\linewidth]{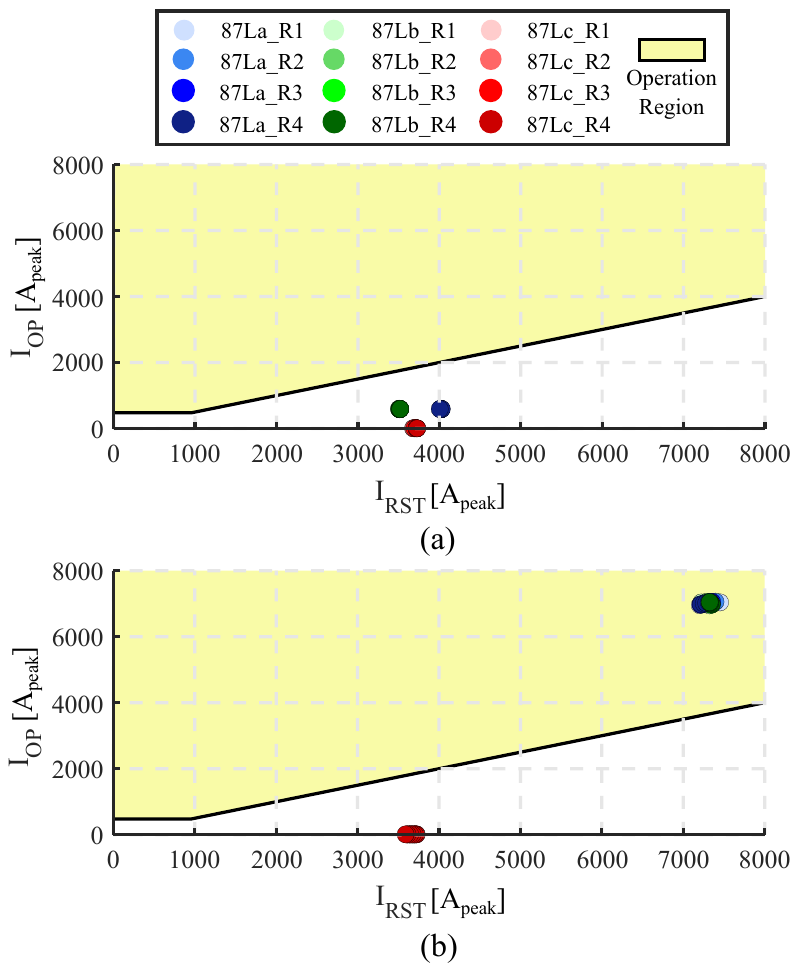}
	\caption{Responses of $87L_{a}$, $87L_{b}$, and $87L_{c}$ elements for AB faults, considering (a) C1 and (b) C2 controls.}\label{fig:Dif-PP-Fase}
\end{figure}
\begin{figure}[!b]
	\centering\includegraphics[width=0.45\linewidth]{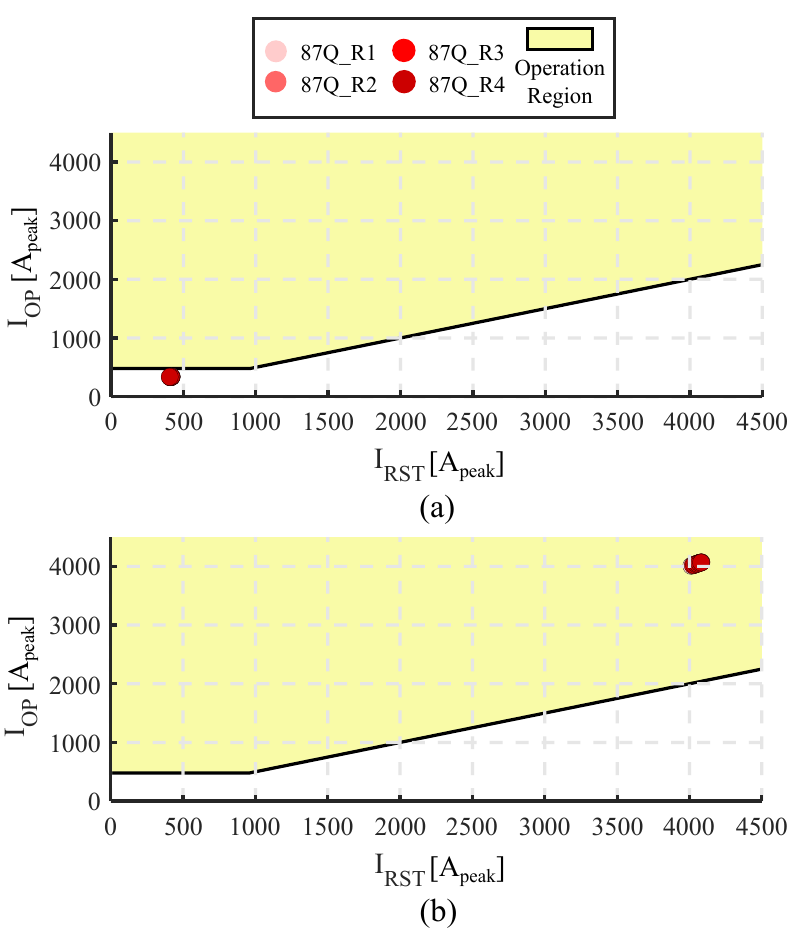}
	\caption{Responses of $87Q$ element for AB faults, considering (a) C1 and (b) C2 controls.}\label{fig:Dif-PP-Seq}
\end{figure}
\begin{figure}[!t]
	\centering\includegraphics[width=0.45\linewidth]{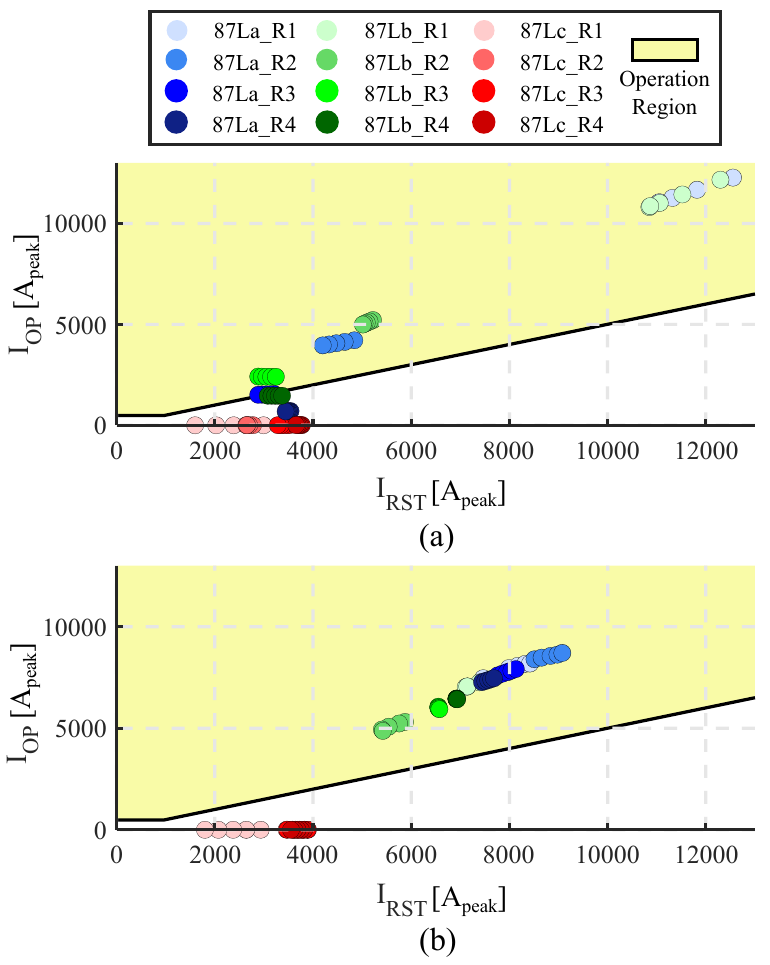}
	\caption{Responses of $87L_{a}$, $87L_{b}$, and $87L_{c}$ elements for ABG faults, considering (a) C1 and (b) C2 controls.}\label{fig:Dif-PPG-Fase}
\end{figure}
\begin{figure}[!t]
	\centering\includegraphics[width=0.45\linewidth]{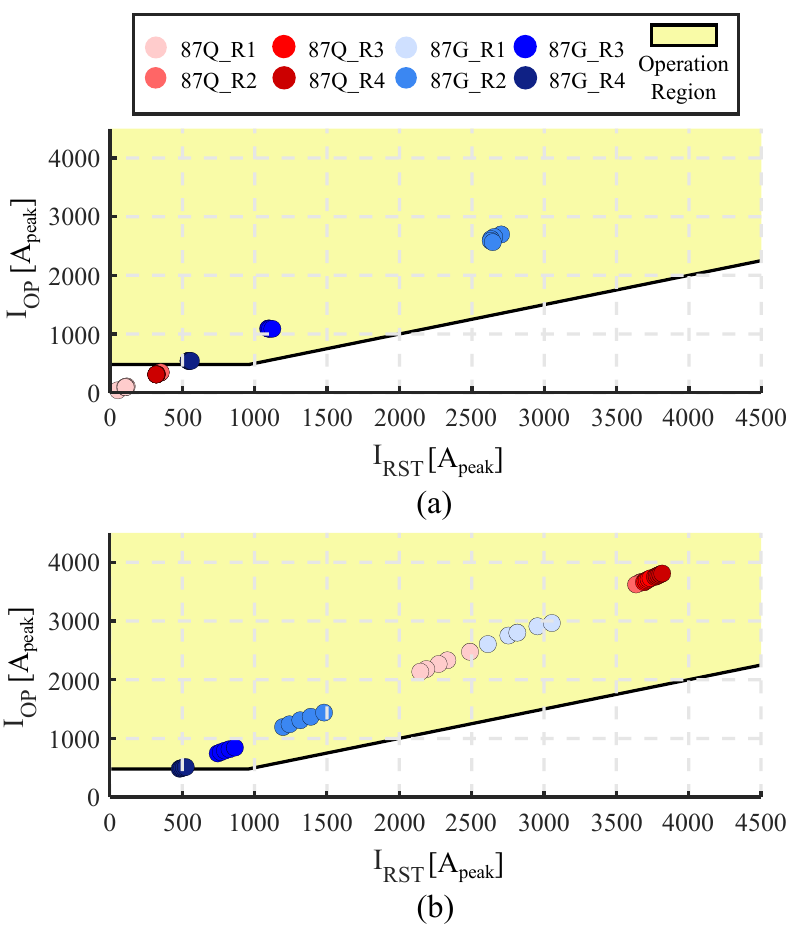}
	\caption{Responses of $87Q$ and $87G$ elements for ABG faults, considering (a) C1 and (b) C2 controls.}\label{fig:Dif-PPG-Seq}
\end{figure}
\begin{figure}[!t]
	\centering\includegraphics[width=0.5\linewidth]{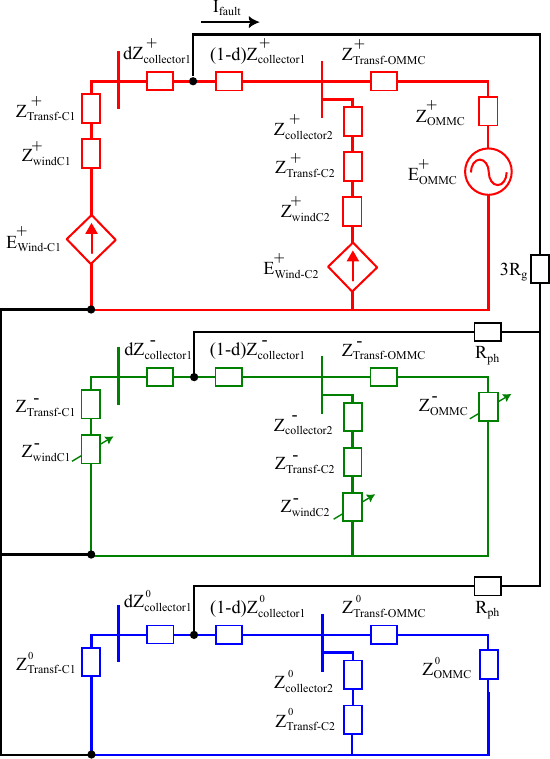}
	\caption{Sequence network for PPG fault.}\label{fig:CircSeqFFT}
\end{figure}

\subsection{Internal Phase-to-Phase (PP) Faults}

For PP faults, Figs. \ref{fig:Dif-PP-Fase} and \ref{fig:Dif-PP-Seq}  illustrate the convergence points of the phase and sequence differential elements, respectively.

The results obtained are very similar to those observed for PG faults with regard to the operation of the differential protections. Since, for this fault type, the positive- and negative-sequence circuits are connected in parallel, nonresidual fault current values will only be obtained when considering C2 controls for the OMMC. In other words, the negative-sequence circuit behaves as an open circuit when both the wind turbines and the OMMC feature negative-sequence current suppression characteristics, preventing the satisfactory operation of the evaluated differential protections.


\subsection{Internal Phase-to-Phase-to-Ground (PPG) Faults}

For PPG faults, Figs. \ref{fig:Dif-PPG-Fase} and \ref{fig:Dif-PPG-Seq}  illustrate the convergence points of the phase and sequence differential elements, respectively.

For this fault type, the sequence network outlined in Fig. \ref{fig:CircSeqFFT} illustrates that the presence of the zero-sequence circuit provides a path for the fault current to be nonzero, even when control C1 is considered for the OMMC. However, it is noted that the fault resistance values define the level of $I_{fault}$, resulting in scenarios where this current does not reach sufficient levels to sensitize all evaluated differential protection elements.

When control C2 is considered for the OMMC, all the differential elements operated satisfactorily, regardless of the fault resistance value considered. Only the 87G element, under $R4$ resistances, approached the limits of the operating region, but still maintained satisfactory operation for internal faults.

On the other hand, considering control C1 for the OMMC, among the evaluated differential elements, it is observed that only the 87G was sensitive for all resistance levels considered. The 87Q element had unsatisfactory performance in all fault scenarios, and regarding the phase elements, satisfactory operations in phases A and B elements were only achieved for fault resistances $R1$ and $R2$.

Finally, it is emphasized that, as observed for PG faults, the satisfactory operation of the differential protection for this fault type also strongly depends on the transformer winding connections, that is, on the zero-sequence circuit.

\begin{figure}[!t]
	\centering\includegraphics[width=0.45\linewidth]{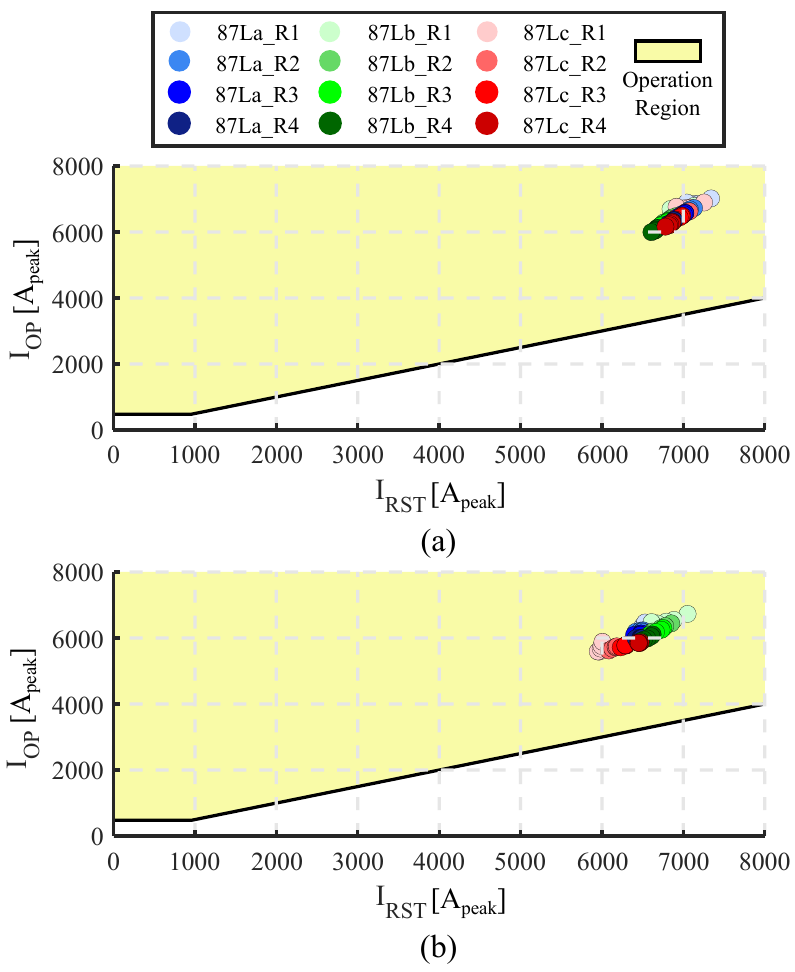}
	\caption{Responses of $87L_{a}$, $87L_{b}$, and $87L_{c}$ elements for ABC faults, considering (a) C1 and (b) C2 controls.}
	\label{fig:Dif-PPP}
\end{figure}
\begin{figure}[!t]
	\centering\includegraphics[width=0.43\linewidth]{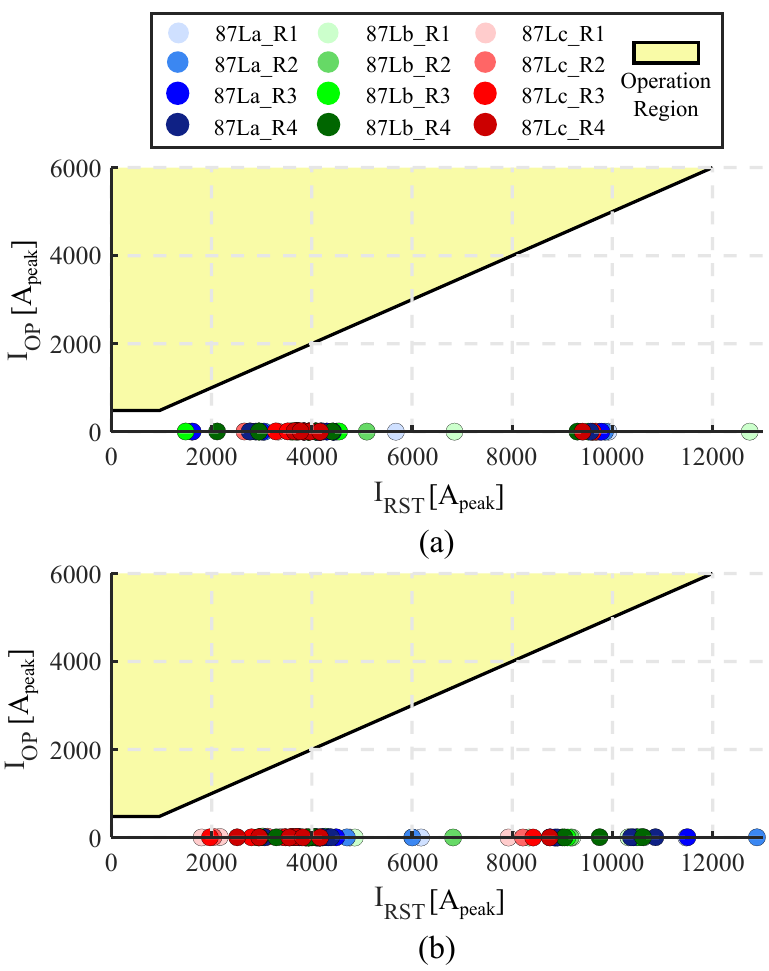}
	\caption{Responses of $87L_{a}$, $87L_{b}$, and $87L_{c}$ elements for external faults, considering (a) C1 and (b) C2 controls.}\label{fig:Dif-Externas-Fase}
\end{figure}
\begin{figure}[!t]
	\centering\includegraphics[width=0.45\linewidth]{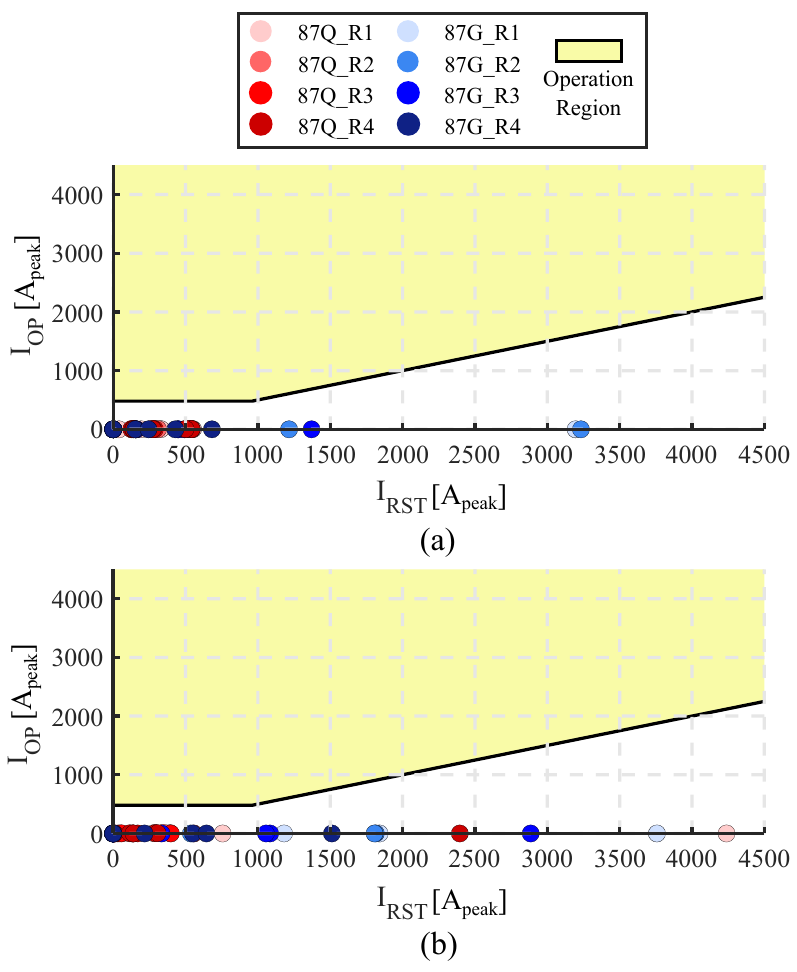}
	\caption{Responses of $87Q$ and $87G$ elements for external faults, considering (a) C1 and (b) C2 controls.}\label{fig:Dif-Externas-Seq}
\end{figure}

\subsection{Internal Three-Phase (PPP) Faults}

For PPP faults, Fig. \ref{fig:Dif-PPP} illustrates the convergence points of the phase differential elements. Since, for this fault type, only the positive-sequence circuit is involved, it was observed that the operation of the assessed phase differential elements was satisfactory for all resistance levels considered and regardless of the OMMC control type (C1 or C2). Thus, for the IBR control strategies assessed in this paper, no situations were identified in which the phase differential protection could be compromised for PPP faults.


\subsection{External Faults}

Finally, faults external to the protected collector cable (at fault points F6 and F7) were simulated to investigate the potential for loss of selectivity in the evaluated differential protection schemes. Figs. \ref{fig:Dif-Externas-Fase} and \ref{fig:Dif-Externas-Seq}  illustrate the convergence points of the phase and sequence differential elements, respectively, considering all assessed fault types and resistances.

As can be observed, there was no situation in which the evaluated IBR control strategies resulted in incorrect operations for external faults, since the operating quantities were practically null regardless of the fault type and resistance considered. However, it is important to note that factors such as synchronization errors and measurement delays were not considered in these studies and could result in nonzero operating currents \cite{BLACK2014}. Nevertheless, since the main focus of this work is to evaluate how IBR control strategies could lead to incorrect operations of differential protections, the assessment of these additional factors was not considered.

\section{Conclusions}

This work comprehensively assessed the performance of conventional and sequence component-based differential protection schemes applied to collector cables in offshore wind farms with MMC-HVDC transmission. The extensive simulation campaign included internal and external faults of varied types and resistances, with detailed electromagnetic transient modeling in PSCAD/EMTDC that reflected realistic converter control strategies and transformer arrangements.

The main findings demonstrate that conventional phase-current-based differential protection schemes may lack sufficient sensitivity to internal faults, especially PG and PP, when both the wind turbines and OMMC employ negative-sequence current suppression strategies. In such scenarios, the negative-sequence circuit effectively opens, suppressing the total fault current and limiting the ability of differential protections to detect internal faults, even at low fault resistances. This reveals a critical vulnerability in traditional protection practices as power systems transition to converter-dominated topologies. The transformer winding configuration, specifically the availability of a zero-sequence path, also proved decisive for the success of differential protection during ground faults.

Enhanced differential protections that incorporate negative- and zero-sequence components (87Q and 87G) improved sensitivity in some fault scenarios. However, similar impacts to those observed for the phase elements were also found in these more sensitive elements for PG and PP faults when considering the C1 control in the OMMC. 

Across the varied fault and control scenarios, no spurious operations were observed due to external faults in the collector cable, underscoring the selectivity of both conventional and sequence-component-based differential schemes. Factors such as synchronization errors and measurement delays were not considered in this paper. 

In conclusion, as renewable integration and converter-dominated grid architectures continue to grow, ensuring sensitive and selective protection for critical collector cables demands innovative approaches beyond conventional relay functions, even when considering differential protections that have historically demonstrated high sensitivity and reliability for protecting systems with IBRs. Thus, the insights from this study are intended to inform both practical protection system design and the ongoing development of future-proofed standards for the next-generation power grids.

\section{Acknowledgements}
The authors thank the Sao Paulo Research Foundation (FAPESP) [$\#$2024/17884-3 and $\#$2025/15469-1] and the  Coordenação de Aperfeiçoamento de Pessoal de Nível Superior – Brasil (CAPES) for their financial support. Additionally, it was funded by FEDER / Ministerio de Ciencia e Innovación - Agencia Estatal de Investigación, under the projects EQUIRED (PID2021-124292OB-I00), HP2C-DT (TED2021-130351BC21) and the Institució Catalana de Recerca i Estudis Avançats (ICREA) Academia Program. We gratefully acknowledge the support of the RCGI – Research Centre for Greenhouse Gas Innovation, hosted by the University of São Paulo (USP), sponsored by FAPESP [$\#$2020/15230-5], and sponsored by TotalEnergies and the strategic importance of the support given by ANP (Brazil’s National Oil, Natural Gas and Biofuels Agency) through the R\&DI levy regulation.

\section*{References}

\bibliographystyle{IEEEtran}
\bibliography{ref}

\end{document}